 \documentclass[final,3p,times,twocolumn]{elsarticle}

\usepackage{float}
\journal{Physics Letters A}

\usepackage{graphicx}
 \usepackage{epsfig}
\begin{document}
\def\ds{\displaystyle}
\begin{frontmatter}

\title{Super Bloch Oscillation in a $\mathcal{P}\mathcal{T}$ symmetric system}
\author{Z. Turker$^{\dagger}$, C. Yuce$^{\star}$ }
\address{ $^{\dagger}$Engineering Faculty, Near East University, North Cyprus, \\
$^{\star}$Physics Department, Anadolu University, Eskisehir, Turkey}
\ead{$^{\star}$cyuce@anadolu.edu.tr} \fntext[label2]{}
\begin{abstract}
Wannier-Stark ladder in a $\mathcal{P}\mathcal{T}$ symmetric system is generally complex that leads to amplified/damped Bloch oscillation. We show that a non-amplified wave packet oscillation with very large amplitude can be realized in a non-Hermitian tight binding lattice if certain conditions are satisfied. We show that pseudo $\mathcal{P}\mathcal{T}$ symmetry guarantees the reality of the quasi energy spectrum in our system.  
\end{abstract}


\end{frontmatter}


\section{Introduction}
  
A quantum particle in a periodical potential driven by a constant force performs an oscillatory motion instead of uniformly accelerated motion. This oscillation is known as Bloch oscillation \cite{blochorjinal}.  Long after its theoretical prediction, it was observed in some physical systems such as superlattice \cite{blochsemi}, ultracold \cite{blochultracold} and optical systems \cite{blochopt}. If the system is subjected to an oscillating force instead of the constant force, then localization occurs at some certain oscillation frequencies.  This kind of localization is known as dynamical localization and its physical origin is different from Anderson localization \cite{DL}. The combined presence of both constant and oscillating forces leads to the photon-assisted tunneling. In this case, wave packet delocalization occurs for an initially localized wave packet when modulation frequency is at multiple integers of the Bloch frequency \cite{pat}. Another interesting effect is the super Bloch oscillations, which arises when an integer multiple of the oscillation frequency is only slightly detuned from the Bloch frequency associated with the constant force \cite{superbloch1,superbloch2,superbloch3,superbloch4,superbloch5,superbloch6}. Note that the amplitude of super Bloch oscillation is much larger than that of Bloch oscillation. It was experimentally realized in an ultracold system, where a giant matter-wave oscillation (super Bloch oscillation) that extend over hundreds of lattice sites was observed \cite{superbloch2}. A generalization of super Bloch oscillation to correlated particles was considered in \cite{superbloch5}. We refer the reader to \cite{superbloch6} for the theoretical analysis of the Hermitian case.\\
The experimental realizations of non-Hermitian $\mathcal{P}\mathcal{T}$
symmetric optical systems \cite{deney1,deney2,deney3} have triggered interests in the investigation of oscillatory dynamics in $\mathcal{P}\mathcal{T}$ symmetric systems. Bloch oscilation in a $\mathcal{P}\mathcal{T}$ symmetric complex crystal was first addressed theoretically in \cite{ptbloch1}. It was shown that either amplified or damped Bloch oscillation occurs depending on the sign of the constant force. This is simply due to broken $\mathcal{P}\mathcal{T}$ symmetry by the external force that leads to the appearance of complex Wannier Stark (WS) ladders. Bloch oscillations were also shown to display unusual features such as non-reciprocal cycles related to a violation of Friedel's law of Bragg scattering and cascade of wave packet splittings. Bloch oscillations for a non-Hermitian infinitely-extended, ring and truncated tight-binding lattices with unidirectional hopping were investigated \cite{ptblochek}. It was shown that the oscillatory motion is not secularly damped nor amplified for that system as a result of the equally-spaced real Wannier-Stark ladder. It is well known that the acceleration theorem explains Bloch oscillation semi-classically for Hermitian lattice. An extension of the the acceleration theorem to non-Hermitian lattices was discussed and Bloch oscillation for the generalized non-Hermitian Hatano and Nelson was specifically studied \cite{ptbloch2}. Another theoretical paper investigated Bloch-Zener oscillation for locally $\mathcal{P}\mathcal{T}$ symmetric system \cite{ptbloch3}. $\mathcal{P}\mathcal{T}$ symmetric Bloch oscillation is not only theoretical interests. It was also experimentally realized in \cite{ptbloch4,ptbloch5,ptbloch10}. Although the non-Hermitian Hamiltonian considered in \cite{ptbloch4} is  $\mathcal{P}\mathcal{T}$ symmetric, the total power was observed to be changed in time. In the experiment \cite{ptbloch5}, pseudo-Hermitian wave packet propagation with a vanishing net increase in power was observed by controlling the period of the Bloch oscillation in a global and local $\mathcal{P}\mathcal{T}$ symmetric mesh lattice.  A resonant restoration of $\mathcal{P}\mathcal{T}$ symmetry and secondary emissions, which occurs each time the wave packet passes through the exceptional point was also observed in that experiment. Another interesting dynamical effect in solid state and optical systems is the dynamic localization effect. The dynamic localization for non-Hermitian $\mathcal{P}\mathcal{T}$ symmetric systems have been investigated by various authors \cite{ptdynloc1,ptdynloc3,ptdynloc4}. It was shown that  the reality of the quasi-energy spectrum is preserved by the $\mathcal{P}\mathcal{T}$ symmetry and  Bragg scattering in the crystal becomes highly nonreciprocal at the $\mathcal{P}\mathcal{T}$ symmetry-breaking point \cite{ptdynloc3}.\\
In this paper, we study another interesting dynamical effect, super Bloch oscillation, for a complex lattice. Bloch oscillation in non-Hermitian lattices is generally either amplified or damped because of complex Wanner-Stark ladders. Here, we show that undamped/non-amplified large amplitude oscillation occurs in a non-Hermitian lattice if certain conditions are satisfied.

\section{Model}

Consider an array of tight-binding one dimensional lattice
with alternating gain and loss. The system is also subject to a time-dependent force. The non-Hermitian Hamiltonian for our system is given by
\begin{eqnarray}\label{ham}
H= -\sum_{n}
T (|n><n+1|+|n+1><n|)\nonumber\\
+\sum_{n}\left(F(t)~a~
n+i(-1)^n\gamma\right) ~|n><n|
\end{eqnarray}
where the constant $\ds{T}$ is tunneling amplitude through
which particles are transferred between neighboring sites, $F(t)$
is a real valued time-dependent force, $a$ is the distance between the centers of adjacent sites and the real constant $\ds{\gamma}$ is
non-Hermitian degree describing the strength of the gain/loss materials. Note that the same Hamiltonian can also describe an array of optical waveguides if the time parameter $t$ is replaced with the propagation distance $z$.  \\
Let us first study the reality of the spectrum of this non-Hermitian Hamiltonian. As a result of the balanced gain and loss, the non-Hermitian Hamiltonian becomes $\mathcal{P}\mathcal{T}$
symmetric if no force exists $F=0$. Therefore the corresponding spectrum is real unless $\gamma$ exceeds a critical value $\gamma_{PT}$, which corresponds to the transition from unbroken to broken  $\mathcal{P}\mathcal{T}$ symmetry. In the presence of force, the reality of the spectrum depends on the form of the force. For example, a time-independent force, $F(t)=const.$, breaks $\mathcal{P}\mathcal{T}$ symmetry and consequently complex-conjugate eigenvalues appear. In other words, Wannier-Stark ladders become complex-valued and then a localized wave packet is amplified/damped during Bloch oscillation. The physics changes significantly if the force changes periodically in time. In this case, the energy spectrum is replaced by a quasi-energy spectrum, which is real for a wide range of parameters \cite{ptdynloc3}. At some certain parameters of the Hamiltonian, dynamic localization was shown to occur. In this paper, we suppose that the force is composed of both a constant and an oscillating terms
\begin{equation}\label{revf241}
F(t)=\omega_0\left(~  l+ \kappa\cos({\omega_0}t+\phi  ) ~ \right)
\end{equation}
where  $\omega_0$ is the modulation
frequency, $\ds{l{\neq}0}$ is an integer, the constant
$\ds{\omega_0 \kappa}$ is the strength of the oscillating
term and $\ds{\phi}$ is the initial phases. Apparently, the constant term in (\ref{revf241}) breaks the $\mathcal{P}\mathcal{T}$
symmetry of the Hamiltonian (1). So, one expects that the system has
zero threshold for $\mathcal{P}\mathcal{T}$ symmetry breaking,
i.e. $\ds{\gamma_{ \mathcal{P}\mathcal{T}}=0}$. This means that wave packet is either amplified or damped during oscillation. However, this is not the case since pseudo $\mathcal{P}\mathcal{T}$ symmetry guarantees the reality of the corresponding spectrum as it was shown in \cite{cempseudo,pseudo2,zhong}. As a result, pseudo $\mathcal{P}\mathcal{T}$ symmetry allows us to study large amplitude oscillation of a wave packet that is neither damped nor amplified.\\
We now show that the spectrum in our system is real. One can use high-frequency Floquet approach to construct a time-independent effective Hamiltonian to study the spectrum of the periodical Hamiltonian. In this approach, which is valid if $\omega_0$ is large, the
tunneling parameter is replaced by an effective tunneling
parameter, $\ds{{T_{eff.}} }$ \cite{superbloch3,effect0,effect1}
\begin{eqnarray}\label{ameffh}
H_{eff.}= -\sum_{n=1}
T_{eff.}|n><n+1|+{T^{\star}_{eff.}}~|n+1><n|\nonumber\\+i\gamma\sum_{n=1}^N
(-1)^n~ |n><n|~~~~
\end{eqnarray}
where star denotes the complex conjugate and the effective
tunneling is given by
\begin{eqnarray}\label{efftun}
\frac{T_{eff.}}{T}=\overline{ \int_0^t e^{i\eta}dt^{\prime}}
\end{eqnarray}
where overline denotes the average over time and $\eta$ is given by
$\ds{\eta(t)=\int_0^{t}{F(t^{\prime})~dt^{\prime}}}$. To evaluate the integral (\ref{efftun}), we use the Jacobi-Anger expansion; $\ds{
e^{i\kappa\sin(x)}=\sum_{m} \mathcal{J}_m ( \kappa )e^{imx} }$,
where $\mathcal{J}_m$ is the
$m$-th order Bessel function of first kind. If we expand the
oscillatory term $\ds{e^{i\eta}}$ in terms of Bessel functions and take the time average of the integration, we get the effective tunneling expression
\begin{eqnarray}\label{son202gh}
\frac{T_{eff.}}{T}= \mathcal{J}_{-l} (\kappa)e^{-il\phi}
\end{eqnarray}
We conclude that the force (\ref{revf241}) modifies the tunneling amplitude in the Hamiltonian. As it was discussed above, this type of force breaks the  $\mathcal{P}\mathcal{T}$ symmetry of the original Hamiltonian. However, this $\mathcal{P}\mathcal{T}$ symmetry breaking term is absent in the effective Hamiltonian. The system is called pseudo $\mathcal{P}\mathcal{T}$ symmetric, which arises when not the  original Hamiltonian but the effective Hamiltonian is $\mathcal{P}\mathcal{T}$ symmetric \cite{cempseudo,pseudo2}. \\
The absolute value of the effective tunneling parameter changes with $\ds{\kappa}$ and $\ds{l}$. The
Bessel function $\ds{ \mathcal{J}_{-l} (\kappa)}$ is roughly like a decaying sine function. In the absence of the oscillating force term, $\kappa=0$, the
Bessel function is always zero, $\ds{ \mathcal{J}_{-l}(0)=0}$.
Therefore, the effective tunneling is suppressed and the spectrum
becomes complex. This result is a direct consequence of the broken $\mathcal{P}\mathcal{T}$ symmetry that occurs
when $l$ changes from zero to nonzero.  Note that vanishing $T_{eff.}$ does not mean entire destruction of tunneling because of the neglected off-resonant terms in the derivation of the effective tunneling. The tunneling is partially
restored and the system enters the pseudo $\mathcal{P}\mathcal{T}$
symmetric phase with the additional application of oscillating force term. This is the region where we study super Bloch oscillation. We emphasize that the pseudo $\mathcal{P}\mathcal{T}$ symmetry is spontaneously broken whenever $\ds{\kappa}$ is a root of Bessel function of order $l$ since the effective tunneling vanishes.\\
Having derived the effective Hamiltonian, we can discuss the reality of the spectrum. For an infinitely extended lattice, the energy eigenvalues of the effective Hamiltonian are composed by two minibands and given by $\ds{E_{eff.}=\mp  \sqrt{ 4 |T_{eff.}|^2 \cos^2(ka)  -\gamma^2}}$. Therefore the spectrum is complex at any $\ds{\gamma}$ for the periodical lattice. If the system, on the other hand, is truncated, the spectrum of the effective system becomes real  provided that non-Hermitian degree is below than a critical number. If it is beyond the critical number, spontaneous
$\mathcal{P}\mathcal{T}$ symmetry breaking occurs and the
eigenfunctions of the Hamiltonian are no longer simultaneous
eigenfunction of $\mathcal{P}\mathcal{T}$ operator. The energy spectrum of the effective Hamiltonian (\ref{ameffh}) for the truncated lattice can be found numerically. In our system, the critical value decreases with increasing number of truncated lattice sites. As $N\rightarrow\infty$, the critical value goes to zero. A discussion and analytical formula for open and periodical boundary conditions can be found in \cite{ekfhksa}.
The above formalism is valid when $t>>1/\omega_0$. It does not account for the dynamics of the system, either. However, it gives an insight to us on the reality of the spectrum. This motivates us to study undamped/non-amplified super Bloch oscillation in our non-Hermitian system. Below, we study the dynamics of the system numerically.

\section{Super Bloch Oscillations}

To observe super Bloch oscillation, the frequency $\ds{\omega_0}$ should be slightly detuned from the Bloch frequency. To satisfy this off-resonance condition, we can make the following replacement in the expression (\ref{revf241}): $\ds{l{\rightarrow} l+\delta}$, where the parameter $\delta<<1$ is a small detuning. The detuning $\delta$ makes the amplitude of oscillation much larger than that of Bloch oscillation as can be seen below. We stress that $\delta$ acts perturbatively to the effective Hamiltonian (\ref{ameffh}) and hence we say that corresponding spectrum changes slightly.\\
Let us first review the super Bloch oscillation briefly for the Hermitian system. The semiclassical explanation for the super Bloch oscillation is as follows. In the absence of gain and loss, $\gamma=0$, the energy expression for the tight binding lattice is given by $\ds{E=-2T\cos(ka)}$, where the quasi-momentum $k$ changes with the force according to $dk/dt=F(t)$. Then the  instantaneous group velocity for the force (\ref{revf241}) reads $\ds{v_g=dE/dk=2T a\sin((l+\delta)\omega t+\kappa\sin(\omega t))}$ (We set $\phi=0$ for simplicity). One can compute the average of $\ds{v_g}$ over one period using the Jacobi-Anger relation. Then the mean position of the wave packet can be found using $\bar{v}_g=d\bar{x}_g/dt$, where overline denotes average
\begin{equation}\label{ptsb1}
\bar{x}_g(t)=2Ta ~\mathcal{J}_{-l}(\kappa) ~ \frac{\cos(\delta \omega_0 t)}{\delta ~\omega_0}
\end{equation}
This expression clearly shows us why the oscillation is called super Bloch oscillation. This is because of the fact that the amplitude of the oscillation scales with the inverse of the detuning $\delta$. The amplitude of the oscillation can be enhanced and oscillation with a giant amplitude can be observed if the detuning is made small enough. Not only the amplitude but also the period of the super Bloch oscillation is large. The higher the amplitude of the oscillation, the larger the period is.   \\
The semi-classical picture for the Hermitian lattice \cite{superbloch6} can not be directly generalized to a non-Hermitian lattice as discussed in \cite{ptblochek}. Below, we perform numerical computation to investigate super Bloch oscillation for the non-Hermitian truncated lattices. We take $T=1$ and $N=30$ from now on. Let us first write the solution as $\ds{|\psi(t)>=\sum_{n=1}^N c_n(t)|n>}$, where $n$ is the lattice site number. The corresponding equation for the time-dependent coefficients $\ds{c_n(t)}$ can be obtained by substituting this expansion into the Schrodinger-like equation, $\ds{H|\psi>=i\dot{\psi}}$. Suppose first that $\gamma=0$. The Fig. 1 depicts the evolution of a wave packet  (snapshot of $|c_n(t)|^2$) for a single site initial input, $c_n(t)=\delta_{n,15}$ for the parameters $l=1$ and $\ds{\delta=0.1}$. For comparison, the intensity profile for Bloch ($\kappa=0$) and super Bloch oscillations ($\kappa=0.5$) are shown in the figure. The presence of the constant force forms the Wannier-Stark ladder and therefore Bloch oscillation occurs as can be seen from the upper panel. The additional presence of the oscillating force together with the detuning $\delta=0.1$ makes the amplitude of this oscillation larger. 
As can be seen from the lower panel, the amplitude and the period of the super Bloch oscillation are larger than those of the Bloch oscillation.\\
\begin{figure}[t]\label{20}
\includegraphics[width=6.0cm]{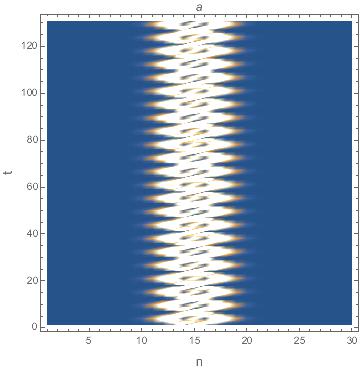}
\includegraphics[width=6.0cm]{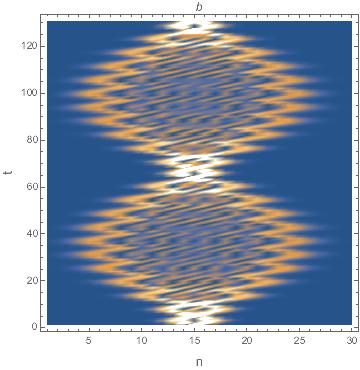}
\caption{ The parameter for both figures are given by $N=30$, $l=1$, $\delta=0.1$ and $\omega_0=1$. In the upper figure, localization occurs when $\kappa=0$ and we see Bloch oscillation. In the lower one, we see large amplitude super Bloch oscillation when $\kappa=0.5$. The horizontal and vertical axes represent $N$ and time, respectively.}
\end{figure}
Let us now study $\mathcal{P}\mathcal{T}$ symmetric super Bloch oscillation, $\gamma\neq0$. In the absence of the oscillating force, $\kappa=0$, the energy spectrum becomes complex and consequently the total intensity grows exponentially in time. As discussed above, the presence of an additional oscillating force makes the effective Hamiltonian $\mathcal{P}\mathcal{T}$ symmetric although the original Hamiltonian is not. Therefore, super Bloch oscillation without amplified intensity can be observed as long as $\gamma$ is below than a critical value, $\gamma_{PT}$. If it exceeds the critical value, amplification/damping occurs. Consider first $\ds{\gamma<\gamma_{PT}}$. In this case, energy spectrum is real and we don't expect exponential growth of the total particle number. We numerically compute time evolution for a single site and a broad Gaussian initial excitations. The figure-2 plots the breathing and oscillatory super Bloch oscillation corresponding to an initial single-site and broad-site excitations of the lattice. As can be seen from the figure, large amplitude oscillation similar to the one in the Hermitian lattice occurs. However, we find that particle number is not conserved but changes slightly with time. This is because of the fact that eigenstates of non-Hermitian Hamiltonian are not orthogonal to each other. A similar effect was also observed in an experiment on Bloch oscillation for a $\mathcal{P}\mathcal{T}$ symmetric Hamiltonian \cite{ptbloch4}. We emphasize that non-conservation of the total particle number breaks the exact self-imaging property of the lattice. However, this effect is weak and the oscillation is almost periodical if $\ds{\gamma<\gamma_{PT}}$. Note that the amplitude of the oscillation can be made higher by choosing the detuning $\ds{\delta}$ smaller than $0.1$. However, our lattice is truncated and once the wave packet reaches the lattice boundary, the oscillating character of the wave packet is lost. To sum up, we show that although (small amplitude) stable Bloch oscillation does not occur in our non-Hermitian lattice, large amplitude Bloch oscillation is possible in the same system. This is the main finding of this paper. Suppose next that the effective system is in the spontaneously broken $\mathcal{P}\mathcal{T}$ symmetric phase, $\gamma>\gamma_{PT}$ , where pairs of complex-conjugate eigenvalues appear. In this case, wave packet is amplified during the oscillation. In fact, the total number of particles grows exponentially at almost every lattice sites even if only single site is initially excited. Therefore, super Bloch oscillation does not occur in this case. If, on the other hand, $\gamma$ is very close to $\gamma_{PT}$, we can still observe at least a few cycle large amplitude oscillation. As time goes on, new particles that appear at every lattice sites become too much and oscillatory motion of the wave packet is consequently destroyed. The Fig-3 shows numerical simulation for this case when the initial wave function is Gaussian. As can be seen, the wave packet undergoes high amplitude oscillations in real space and secondary emission occurs. \\
 \begin{figure}[t]\label{20}
\includegraphics[width=6.0cm]{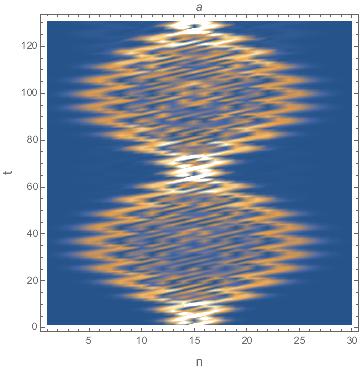}
\includegraphics[width=6.0cm]{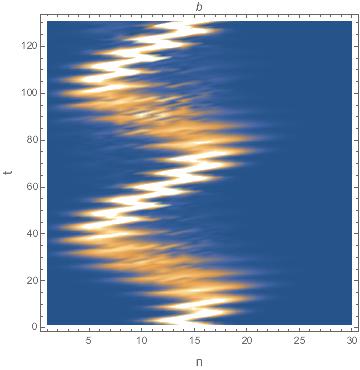}
\caption{ The parameters are the same as in the figure 1, but $\gamma=0.05$ ($\gamma_{PT}=0.1$)
If only single site in initially populated, breathing super Bloch oscillations occurs (a). In the case of broad Gaussian initial excitation, the wave packet undergoes an oscillatory motion (b). The horizontal and vertical axes represent $N$ and time, respectively. Note that super Bloch oscillation is destroyed after a couple of cycles that corresponds to a large propagation distance. }
\end{figure}
\begin{figure}[t]\label{20}
\includegraphics[width=6.0cm]{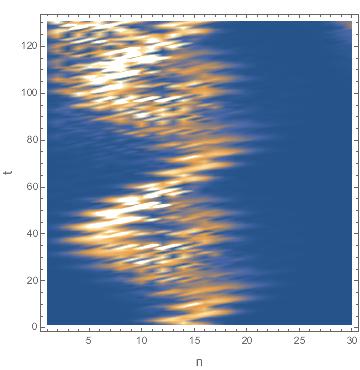}
\caption{ The parameters are the same as in the figure 1, but $\gamma=0.25$ for a broad initial wave.}
\end{figure}
So far, we have only considered super Bloch oscillation. For our problem, dynamic localization and photon assisted tunneling in the $\mathcal{P}\mathcal{T}$ symmetric region can also be investigated. Let us mention them briefly. The dynamic localization appears in the presence of the oscillating force, $\ds{l=0, \kappa\neq0}$. At some particular strength of the oscillating force, an initially localized wave-packet stops spreading in time. In other words, the system is in the dynamically localized regime if the effective tunneling vanishes, which occurs whenever $\kappa$ is a root of the Bessel function of order $\ds{l}$. Note that the phase $\phi$ plays no role on dynamic localization. If the effective tunneling is zero, then the critical point $\ds{\gamma_{PT}}$ is zero, too. We numerically solve our equation for $\ds{l=0}$, $\ds{\kappa=2.40}$. (Note that $\ds{\mathcal{J}_0(2.40)=0}$). We see that the total number of particles grow exponentially just as $\gamma$ becomes different from zero. Secondly, let us mention about photon assisted tunneling. It is well known that the translational symmetry of the lattice is broken and tunneling is consequently suppressed by the presence of the constant force. Thus, an initially localized wave-packet remains localized in the Hermitian lattice. However, tunneling is partially restored if an additional oscillating force is also present in the system. This effect is known as photon assisted tunneling \cite{superbloch3} since the role of the photons is played by the oscillation of the force at resonant frequencies. In our system, partial restoration of the tunneling increases $\ds{\gamma_{PT}}$. In the presence of only constant force, $\ds{\gamma_{PT}=0}$ and the corresponding Wannier-Stark ladder are complex, while the additional oscillating force switches $\ds{\gamma_{PT}}$  from zero value to a non-zero value. \\
In conclusion, the main finding of our paper is the existence of non-amplified super Bloch oscillation in a $\mathcal{P}\mathcal{T}$ symmetric lattice even though Bloch oscillation is amplified in the same lattice. In a non-Hermitian lattice, the standard semi-classical picture does not work. Using the high-frequency
Floquet analysis, we have shown that super Bloch oscillation can occur if the non-Hermitian system is in pseudo $\mathcal{P}\mathcal{T}$ symmetric region. We have numerically seen that the super Bloch oscillation in our system with $\ds{\gamma<\gamma_{PT}}$ is very similar to the one in the Hermitian lattice, $\gamma=0$ with the exception that the particle number changes little because of the nonorthogonality of eigenvectors. We have discussed that super Bloch oscillation in our non-Hermitian tight binding lattice is destroyed if $\ds{\gamma>\gamma_{PT}}$. On the other hand, we have shown that a few cycle super Bloch oscillation with secondary emission that has no analogue in a Hermitian system can be observed if $\ds{\gamma}$ is very close to $\ds{\gamma_{PT}}$. We have also discussed dynamic localization and photon assisted tunneling in our non-Hermitian system.

\end{document}